\documentclass[eqsecnum,showkeys,noshowpacs,nofootinbib,aps,epsfig]{revtex4}
\renewcommand{\theequation}{\arabic{equation}}
\usepackage[pdftex]{graphicx}
\def\bea{\begin{eqnarray}}
\def\eea{\end{eqnarray}}

\newcommand{\nn}{\nonumber}

\def\beq{\begin{equation}}
\def\eeq{\end{equation}}

\def\pa{\partial}

\newbox\pippobox

\begin{document}
\input epsf
\title{Photon size in higher dimensional phantom cosmology}
\author{Soon-Tae Hong}
\email{galaxy.mass@gmail.com}
\affiliation{Center for Quantum Spacetime and Department of Physics, Sogang University, Seoul 04107, Korea}
\date{\today}
\begin{abstract}
We study a higher dimensional cosmology with phantom field associated with a negative kinetic term. 
Assuming that the universe possesses the phantom field defined in $D$ dimensional spacetime, we investigate in detail 
the solutions involved in the higher dimensional phantom cosmology, to explicitly predict photon size and phantom field strength at present 
in nature. To be specific, we find that the photon size 
decreases drastically at the early stage of the universe after the Big Bang. Next we explicitly demonstrate the dependences of the 
photon size, universe size and phantom field strength on the spacetime dimensionality $D$. We observe that the size of the universe 
undergoes stiff explosion with different types of slope depending on $D$. Moreover the scale factor of the universe at present is shown to 
approach to a saturated value, which is independent of $D$ and is the same as that in the $D=4$ Friedmann-Robertson-Walker cosmology. The photon size and phantom field strength in the greater dimensionality are also shown to be larger and lower than those in the smaller one, 
respectively. Next the photon size at present $b_{*}$  in $D=5$ is numerically shown to be extremely small, namely $b_{*}=6.08\times 10^{-216}$ cm, comparing to $b_{*}=1.56\times 10^{-63}$ cm in $D=10$. In contrast, the phantom field strength at present $\sigma_{*}$ is shown to be relatively large $\sigma_{*}=4.72\times 10^{24}$ (dyne$)^{1/2}$ in $D=5$, comparing to $\sigma_{*}=1.39\times 10^{22}$ (dyne$)^{1/2}$ in $D=10$.
\end{abstract}
\pacs{14.70.Bh, 11.25.-w, 04.60.Cf, 04.50.-h, 11.10.Kk, 98.80.-k, 98.80.Es, 04.20.Jb}
\keywords{photon size; higher dimensional cosmology; phantom field; universe size}
\maketitle

\section{Introduction}
\setcounter{equation}{0}
\renewcommand{\theequation}{\arabic{section}.\arabic{equation}}

As it is well known, a particle in the string theory is supposed to be an extended object~\cite{witten87,polchinski98}.
Via the string version of the Hawking-Penrose singularity theorem (HPST)~\cite{hawking70}, the stringy cosmology in $D$ dimensional total 
spacetime was investigated~\cite{hong11,hong112} 
with a success that, with the extended string particle, one describes precisely the motion types of stringy congruence in terms of the universe expansion rate after the Big Bang. Moreover, in the stringy HPST, one has an advantage 
that the degrees of freedom of the rotation and shear of stringy congruence are introduced naturally in the early universe. Next in phantom 
cosmology~\cite{caldwell02,lidsey04,faraoni05,hong08,chimento08,kaeonikhom11,wei1111,cannata11,
yang12,astashenok12,novosyadlyj12}, the universe expansion up to the age of the universe has been studied with various phantom field potentials. In particular, 
a higher dimensional phantom cosmology (HDPC) has been exploited to investigate an exact cosmological solution for 
the scale factor of the universe~\cite{hong08}. In the HDPC, the universe has been 
shown to undergo a continuous transition from deceleration to acceleration at some finite time. 
This transition time could be proposed as recent acceleration of the universe. Moreover, a phantom field with an exponential 
potential has been considered to study a solution for the phantom cosmology together with the Hubble parameter~\cite{cannata11}. 
Next the inflationary Big Bang cosmology has been developed into a precision astrophysics by recent cosmological observations 
such as cosmic microwave radiation~\cite{wmap} and supernova data~\cite{perl,riess}. The observations have drawn astrophysics community 
attention to the origin of dark energy~\cite{dark1,dark2,dark3,dark4,dark5}. The phantom cosmology model is also one of the approaches for 
investigating the dark energy~\cite{caldwell02,dark4,dark5}.

The idea of extra dimensions, tracing back to the pioneering works of Kaluza and Klein, has been exploited in various higher dimensional models 
such as the string theory~\cite{witten87,polchinski98}, the Randall-Sundrum cosmology~\cite{randall}, the HDPC~\cite{hong08}, 
the stringy HPST~\cite{hong11,hong112} and the higher dimensional electroweak model~\cite{bolokhov} for instance. Motivated by 
the idea, in this paper 
we will consider the HDPC in $D$ dimensional total spacetime, in order to investigate explicitly the photon size and phantom field strength 
at present and their dependences on $D$. In addition, in $D$ dimensions we will discuss the evolution of the universe size from the Big Bang 
to the present epoch, to demonstrate the dependence of its size on the spacetime dimensionality $D$.

The paper is organized as follows. In Sec. II, we will present the set up of the HDPC with the dimensionality $D$, to study in detail 
the solutions for the differential equations associated with the HDPC. In Sec. III, 
we will investigate phenomenology in the HDPC. Explicitly we will evaluate the photon size at present in the $D$ dimensional 
spacetime. Next we will study the universe and phantom field evolutions from the Big Bang to the epoch at present. 
Sec. IV includes conclusions.  

\section{Set up of HDPC formalism in $D$ dimensions}
\setcounter{equation}{0}
\renewcommand{\theequation}{\arabic{section}.\arabic{equation}}

In this section, we formulate the solutions for the HDPC where the extra dimensions are introduced 
to allow the internal manifold. The HDPC in $D$ dimensional total spacetime is now described by the Lagrangian of the form~\cite{hong08}
\beq 
{\cal L}=\sqrt{-g}\left(\frac{1}{2}R+\frac{1}{2}\alpha g^{MN}\pa_{M}^{~}\sigma\pa_{N}^{~}\sigma
-\Lambda\right),
\eeq 
where $\sigma$ is the phantom field and $\alpha$ is shorthand defined as $\alpha=8\pi G/c^{4}$. Here $M$ and $N$ are 
$D$ dimensional indices and $\Lambda$ is the cosmological constant. In the HDPC with the $D$ dimensional total spacetime, we have 
$(D-4)$ dimensional internal extra manifold, as in the string theory. Note that the string as an extended object is described 
in this internal extra manifold. Inspired by this, in the HDPC the extra manifold will be treated as 
that in which the photon is also described. In this way, we can have the string-like photon.

Variation of the Lagrangian with respect to the 
metric yields the Einstein equations
\beq
R_{MN}=\alpha T_{MN}^{~}-\frac{1}{D-2}\alpha g_{MN}^{~}T+\frac{2}{D-2}g_{MN}^{~}\Lambda,
\label{rmn}\eeq
where the energy-stress tensor is given by
\beq
T_{MN}^{~}=-\pa_{M}^{~}\sigma\pa_{N}^{~}\sigma+\frac{1}{2}g_{MN}^{~}\pa_{Q}^{~}\sigma\pa^{Q}\sigma.
\eeq
We take an ansatz for the metric
\beq
g_{MN}^{~}={\rm diag}~(-1, a^{2}(t), b^{2}(t)), 
\eeq 
where $a(t)$ is the scale factor of the three dimensional universe and  $b(t)$ is the scale factor of the photon described in the
extra $(D-4)$ dimensions. Note that the extra dimensions are associated with the internal torus space~\cite{torus}.

Making use of (\ref{rmn}), we find differential equations with the overdots denoting the derivatives with respect to time,
\bea
3\frac{\ddot{a}}{a}+(D-4)\frac{\ddot{b}}{b}&=&\frac{2}{D-2}\Lambda+\alpha\dot{\sigma}^{2},\nn\\
\frac{\ddot{a}}{a}+2\frac{\dot{a}^{2}}{a^{2}}+(D-4)\frac{\dot{a}}{a}\cdot\frac{\dot{b}}{b}&=&\frac{2}{D-2}\Lambda,\nn\\
\frac{\ddot{b}}{b}+(D-5)\frac{\dot{b}^{2}}{b^{2}}+3\frac{\dot{a}}{a}\cdot\frac{\dot{b}}{b}&=&\frac{2}{D-2}\Lambda,\nn\\
\ddot{\sigma}+\left(3\frac{\dot{a}}{a}+(D-4)\frac{\dot{b}}{b}\right)\dot{\sigma}&=&0,
\label{eoms}
\eea
whose solutions are given by
\bea 
a(t)&=&a_{0}~{\rm exp}~\left[\left(\frac{2}{(D-1)(D-2)}\right)^{1/2}\sqrt{\Lambda}ct
+\frac{n(D-4)}{D-1}\left(\frac{D-2}{2(D-1)}\right)^{1/2}\left(1-{\rm exp}\left[-\left(\frac{2(D-1)}{D-2}\right)^{1/2}
\sqrt{\Lambda}ct\right]\right)\right],\nn\\
b(t)&=&b_{0}~{\rm exp}~\left[\left(\frac{2}{(D-1)(D-2)}\right)^{1/2}\sqrt{\Lambda}ct
-\frac{3n}{D-1}\left(\frac{D-2}{2(D-1)}\right)^{1/2}\left(1-{\rm exp}\left[-\left(\frac{2(D-1)}{D-2}\right)^{1/2}
\sqrt{\Lambda}ct\right]\right)\right],\nn\\
\sigma(t)&=&\sigma_{0}~{\rm exp}~\left[-\left(\frac{2(D-1)}{D-2}\right)^{1/2}\sqrt{\Lambda}ct\right].
\label{bb0}
\eea
Here $a_{0}$, $b_{0}$ and $\sigma_{0}$ are the initial values of $a$, $b$ and $\sigma$ at
$t=0$, respectively, and $n$ is a model parameter to be fixed later. Explicitly, $\sigma_{0}$ is given by
\beq
\sigma_{0}=\frac{n}{D-1}\left(\frac{3(D-2)(D-4)}{2\alpha}\right)^{1/2}.
\label{sigma00}
\eeq
Note that for the case of $D=10$ the above formalism becomes the phantom cosmology in Ref.~\cite{hong08}.

\section{Phenomenology in HDPC}
\setcounter{equation}{0}
\renewcommand{\theequation}{\arabic{section}.\arabic{equation}}

In this section, we proceed to study further novel stringy photon phenomenology using the HDPC. To do this, 
we take an ansatz that, after the Planck time $t_{Planck}=5.39\times 10^{-44}$ second near the Big Bang, the sizes of 
the universe and photon are the same each other so that one can choose $a_{0}$ and $b_{0}$ to become the Planck length~\cite{carroll04}
\beq
a_{0}=b_{0}=l_{Planck}=1.62\times 10^{-33}~{\rm cm}.
\label{a0b0}
\eeq
We exploit the relation 
\bea
\Lambda&=&\frac{(D-1)(D-2)}{6}\Lambda_{*},\nn\\
\Lambda_{*}&=&2.07\times 10^{-56}~{\rm cm}^{-2},
\label{lstar}
\eea
where $\Lambda_{*}$ is the four dimensional cosmological constant. We next define a dimensionless variable $x$ $(0\le x\le 1)$ as 
\bea
t&=&xt_{*},\nn\\
t_{*}&=&4.28\times 10^{17}~{\rm sec},\label{tstar}
\eea
with $t_{*}$ being the age of the universe~\cite{hartnett}.
Inserting $\Lambda$ in (\ref{lstar}) and $t$ in (\ref{tstar}) into (\ref{bb0}), we obtain
\bea
a(x)&=&a_{0}~{\rm exp}~\left[1.07x+\frac{n(D-4)}{D-1}\left(\frac{D-2}{2(D-1)}\right)^{1/2}\left(1-{\rm exp}~[-1.07x(D-1)]\right)\right],
\label{aa02}\\
b(x)&=&b_{0}~{\rm exp}~\left[1.07x-\frac{3n}{D-1}\left(\frac{D-2}{2(D-1)}\right)^{1/2}\left(1-{\rm exp}~[-1.07x(D-1)]\right)\right],
\label{bb02}\\
\sigma(x)&=&\sigma_{0}~{\rm exp}~[-1.07x(D-1)].\label{ss02}
\eea

Making use of the size of the present universe at $x=1$~\cite{davis05}
\beq
a_{*}=4.35\times 10^{28}~{\rm cm},
\label{apresent}
\eeq
and (\ref{aa02}), we fix the value of $n$ to be 
\beq
n=140.37\times\frac{D-1}{D-4}\left(\frac{2(D-1)}{D-2}\right)^{1/2}.
\label{nvalue}
\eeq 
With the above value of $n$, we obtain 
\bea
a(x)&=&a_{0}~{\rm exp}~\left[1.07x+140.37\left(1-{\rm exp}~\left[-1.07x(D-1)\right]\right)\right],
\label{aa03}\\
b(x)&=&b_{0}~{\rm exp}~\left[1.07x-\frac{421.12}{D-4}\left(1-{\rm exp}~\left[-1.07x(D-1)\right]\right)\right],
\label{bb03}
\eea
where $a(x)$ and $b(x)$ describe the evolutions of the universe and photon sizes, respectively.\footnote{Note that in the limit of $D=4$ 
with $b=0$, solving the differential equations in (\ref{eoms}) yields $a=a_{0}~e^{\sqrt{\lambda_{*}/3}ct}=a_{0}~e^{1.07x}$, and 
$\dot{\sigma}=0$ meaning that we do not have a nontrivial solution for the phantom 
field in $D=4$~\cite{hong08}. Moreover in (\ref{nvalue}) we need to have the condition $D\neq 4$ in order 
to obtain a reasonable value of $n$.}

\begin{figure}[!t]
\begin{center}
\includegraphics[width=6.0cm]{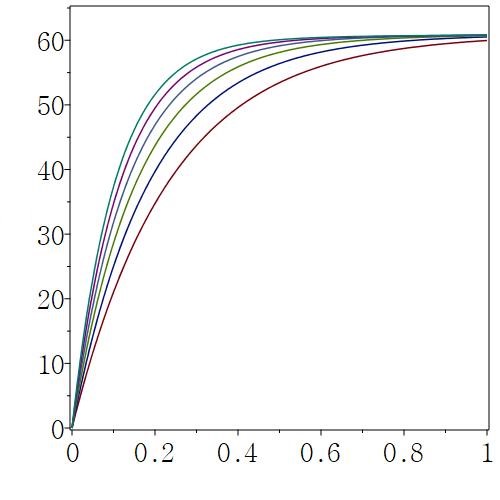}
\includegraphics[width=6.0cm]{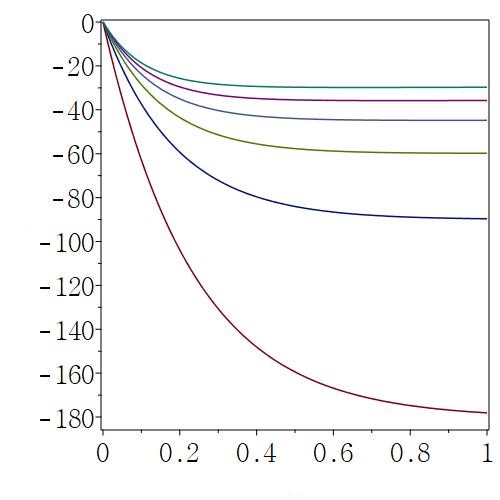}\\
\end{center}
\vskip -0.5cm 
\caption[fig1] {(a) ${\rm log}_{10}[a(x)/a_{0}]$ and (b) ${\rm log}_{10}[b(x)/b_{0}]$ in terms of the dimensionless time variable
$x$. The curves in (a) and (b) are for $D=5,6,7,8,9,10$ from bottom to top.} \label{fig1}
\end{figure}

Now we depict ${\log}_{10}[a(x)/a_{0}]$ in Fig. 1(a) and ${\log}_{10}[b(x)/b_{0}]$ in Fig. 1(b) in terms of the dimensionless time 
variable $x$ $(0\le x\le 1)$. Here we treat the total dimensionality $D$ as a parameter of the curves, to investigate the dependence 
of the scale factors $a(x)$ and $b(x)$ on $D$. Fig. 1(a) shows that the curve is rising up with extreme slope at the early stage of the universe evolution. Note that, along the time evolution, the universe size in the greater dimensionality increases much stiffly than that in the smaller one. In Fig. 1(b) we find that the photon size is drastically falling down at the early stage of the photon evolution after the Big Bang. 
Note that the photon size curve in the greater dimensionality is much higher than that in the smaller one. Next we depict 
${\log}_{10}[\sigma(x)\sqrt{\alpha}]$ in Fig. 2 in terms of the dimensionless time variable $x$. Note that 
all the curves in Fig. 2 decrease linearly along the time evolution. Moreover the dimensionless phantom field in logarithmic scale 
${\log}_{10}[\sigma(x)\sqrt{\alpha}]$ in the greater dimensionality is lower than that in the smaller one.

Now it seems appropriate to comment on the phenomenology of the HDPC {\it at present}. First, putting $x=1$ in (\ref{aa03}) we are 
left with the universe size $a_{*}$ at present
\beq
a_{*}=a_{0}~{\rm exp}~\left[1.07+140.37\left(1-{\rm exp}~\left[-1.07(D-1)\right]\right)\right],
\label{bb042}
\eeq
where $a_{0}$ is now given by (\ref{a0b0}). Note that according to Fig. 1(a) and (\ref{bb042}), there are no significant differences among the universe sizes. In other words, all the curves of the universe sizes converge to a saturated 
value given by (\ref{apresent}), independent of $D$, and the value is the same as that in the $D=4$ Friedmann-Robertson-Walker (FRW) cosmology. This phenomenon is due to the fact that the last term in (\ref{bb042}) is predominant in these curves. 
 
Second, inserting $x=1$ into (\ref{bb03}), we arrive at the photon size $b_{*}$ at present
\beq
b_{*}=b_{0}~{\rm exp}~\left[1.07-\frac{421.12}{D-4}\left(1-{\rm exp}~\left[-1.07(D-1)\right]\right)\right],
\label{bb04}
\eeq
where $b_{0}$ is given by (\ref{a0b0}). Note that at the present epoch there exist very significant differences among the photon sizes as shown in Fig. 1(b). This phenomenon originates from the fact that the last term in (\ref{bb04}) is predominant in these curves. The predicted values of the sizes of the stringy (not point) photon $b_{*}$ in (\ref{bb04}) are listed in Table~\ref{table1} in term of the total dimensionality $D$. In particular, the photon size prediction in $D=10$ case seems to be interpreted as the string size 
itself~\cite{witten87,polchinski98}. Note also that, in the limit of $D=4$, (\ref{bb04}) yields 
$b_{*}=0$ consistent with the fact that we have no extra dimensions in that limit. 

Third, inserting $x=1$ and (\ref{sigma00}) into (\ref{ss02}), we find the phantom field strength $\sigma_{*}$ at present
\beq
\sigma_{*}=\frac{140.37}{\sqrt{\alpha}}\left(\frac{3(D-1)}{D-4}\right)^{1/2}{\rm exp}~\left[-1.07(D-1)\right],
\label{sigmastar}
\eeq
with $\frac{1}{\sqrt{\alpha}}=6.95\times 10^{23}$ (dyne$)^{1/2}$ and $D\neq 4$. The predicted values of $\sigma_{*}$ in (\ref{sigmastar}) 
are listed in Table~\ref{table1} in term of the total dimensionality $D$.\footnote{ Note that the unit of $\sigma^{2}$ is dyne, 
and this phantom field strength seems to be interpreted as a source of detonation of the universe at the early stage of the time evolution 
as shown in Fig. 2. Moreover the phantom field also plays a role of late-time expansion in any epoch of the universe evolution.}

\begin{figure}[!t]
\begin{center}
\includegraphics[width=6.0cm]{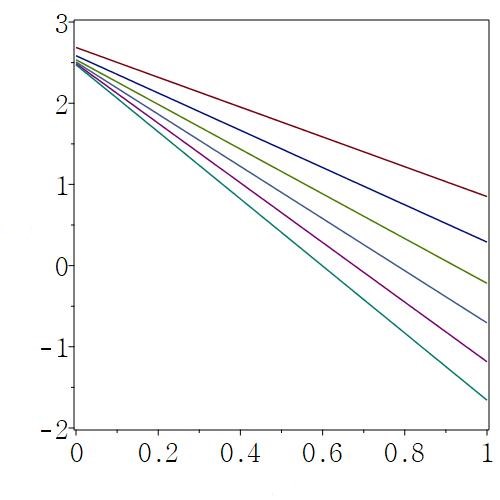}
\end{center}
\vskip -0.5cm 
\caption[fig2] {${\rm log}_{10}[\sigma(x)\sqrt{\alpha}]$ in terms of the dimensionless time variable
$x$. The curves are for $D=5,6,7,8,9,10$ from top to bottom.} \label{fig2}
\end{figure}

\begin{table*}
\caption{The present photon size $b_{*}$ in unit of cm, and the present phantom field strength $\sigma_{*}$ 
in unit of (dyne$)^{1/2}$ in the total dimensionality $D$.} 
\vskip 0.7cm
\begin{center}
\begin{tabular}{ccc||ccc||ccc} 
\hline
$D$ &$~~b_{*}$ &$~~\sigma_{*}$ &$D$ &$~~b_{*}$ &$~~\sigma_{*}$ &$D$ &$~~b_{*}$ &$~~\sigma_{*}$\\
\hline
5 & $~~6.08\times 10^{-216}$ &$~~4.72\times 10^{24}$ &7 & $~~5.12\times 10^{-94}$ &$~~3.95\times 10^{23}$ &9 &$~~1.25\times 10^{-69}$ 
  &$~~4.77\times 10^{22}$\\
6 & $~~1.69\times 10^{-124}$ &$~~1.28\times 10^{24}$ &8 & $~~8.93\times 10^{-79}$ &$~~1.27\times 10^{23}$ &10 & $~~1.56\times 10^{-63}$ 
  &$~~1.39\times 10^{22}$\\
\hline
\end{tabular}
\vskip 0.7cm
\end{center}
 \label{table1}
\end{table*}

According to the stringy HPST~\cite{hong11,hong112}, the photon has a spin structure, due to 
its rotation whose magnitude is assumed to be the same as that of the rotating universe itself. Note that the universe consists of 
celestial objects such as galaxies, stars and planets. If we combine 
the results on the photon rotation from the stringy HPST with those of the HDPC, the photon has an extremely tiny 
size with the spin degree of freedom. The explanation about the photon spin phenomenology seems to be consistent with the corresponding 
experiment.

\section{Conclusions}
\setcounter{equation}{0}
\renewcommand{\theequation}{\arabic{section}.\arabic{equation}}

In summary, we have assumed that the universe has the phantom field associated with a negative kinetic term in the $D$ dimensional spacetime, 
to find the solutions for the differential equations related with the HDPC. Making use of the solutions, we have explicitly evaluated the 
photon size and phantom field strength at present in nature. The formula for the scale factor of the universe has indicated that the 
universe size increases rapidly with different types of slope depending on $D$ at the early stage of the evolution, but after the universe age 
$t_{*}=4.28\times 10^{17}~{\rm sec}$ it approaches to the saturated value independent of $D$, which is the same as that in the $D=4$ FRW cosmology. 
In contrast the photon size has decreased drastically at the early stage of the evolution after the Big Bang, and then the photon size at present $b_{*}$  
in $D=5$ has been shown to be extremely small, comparing to $b_{*}$ in $D=10$, as shown in Table~\ref{table1}. 
We also found that the phantom field strength decreases along the time evolution, and the strength in the greater 
dimensionality is lower than that in the smaller one. The phantom field strengths at present have been numerically predicted in terms of 
the dimensionality as shown in Table~\ref{table1}. Moreover, we have given some comments on the photon spin in 
the framework of the stringy HPST combined with the HDPC. Finally, it is interesting 
to note that the effective equation of state originated from the phantom field is consistent with that 
of the cosmic fluid of the present superacceleration era~\cite{faraoni02,uzan00,faraoni11,hong08}.

\acknowledgments{The author would like to thank J. Lee, T.H. Lee and P. Oh for drawing his attention to the formalism 
of higher dimensional phantom cosmomology. He was supported by Basic Science Research Program through 
the National Research Foundation of Korea funded by the Ministry of Education, NRF-2019R1I1A1A01058449.}

\end{document}